\documentclass[conference]{IEEEtran}
\IEEEoverridecommandlockouts
\usepackage{cite}
\usepackage{amsmath}
\usepackage{amssymb}
\usepackage{amsmath,amssymb,amsfonts}
\usepackage[linesnumbered, ruled,commentsnumbered]{algorithm2e}
\usepackage{algorithmic}
\usepackage{graphicx}
\usepackage{textcomp}
\usepackage{soul}
\usepackage{xcolor}
\usepackage{subfigure}  %调用子图
\usepackage{booktabs}  %表格加粗
\usepackage{tikz}
\usepackage{pgfplots}

\usetikzlibrary{positioning, shapes.geometric}
\def\BibTeX{{\rm B\kern-.05em{\sc i\kern-.025em b}\kern-.08em
    T\kern-.1667em\lower.7ex\hbox{E}\kern-.125emX}}

\begin{document}

\title{Path Planning Algorithm Comparison Analysis for Wireless AUVs Energy Sharing System \\
\thanks{This work was supported by National Science Foundation of Top Talent of SZTU (No. GDRC202210), Shenzhen Technology University Self-made Laboratory Equipment Project (No.JSZZ202301008)}
}

\author{\IEEEauthorblockN{Zhengji Feng\textsuperscript{}, Hengxiang Chen\textsuperscript{}, Heyan Li\textsuperscript{}, Xiaolin Mou\textsuperscript{*}}
	\IEEEauthorblockA{\textit{College of Urban Transportation and Logistics, Shenzhen Technology University} \\
		mouxiaolin@sztu.edu.cn}
}

\maketitle

\begin{abstract}
Autonomous underwater vehicles (AUVs) are increasingly used in marine studies, military applications, and undersea exploration. However, the battery performance affects
AUVs range. In this paper, a framework of wireless AUVs energy sharing system is proposed which can do rapidly energy replenishment for AUVs. The path planning is a crucial aspect in
AUVs energy sharing and automation, which requires to generate trajectories to specified goals. This article mainly focuses on avoiding irregular obstacles efficiently and narrow area passing ability at underwater environment based on Rapidly-exploring Random Trees Star (RRT*) and Particle Swarm Optimisation (PSO). The comparison and analysis of the two algorithms through the simulation results in the random obstacle environment and the irregular obstacle environment are provided in the finally.
\end{abstract}

\begin{IEEEkeywords}
AUVs, Path Planning, Irregular obstacles, Narrow area, RRT*, PSO
\end{IEEEkeywords}

\section{Introduction}
As the development of autonomous underwater vehicles(AUVs), more and more sensors are being equipped, such as cameras, scanning sonar, inertial measurement units, \emph{etc.}. Meanwhile, it is being used in wide range of military and agricultural application, such as undersea exploration, hydrological surveys \cite{T_Emre,Z_Liu,T_Ken} \emph{etc.}. However, the endurance of AUVs is limited by the onboard energy storage among which the battery systems dominate. AUVs usually use the battery swapping and physical power cable (conductive charging), and these methods are not convenient. Another method is the AUVs docking station. This system should adapt the accurate navigation system with the expensive price \cite{Y_Lei}. Wireless charging technology has far-reaching implications for improving the cruising capabilities of AUVs underwater. 

A prototype wireless AUVs charger with a funnel structure is proposed by Agostinho \emph{et al.}, aiming to integrate commercially available IPT technologies while allowing for most AUVs to be used. It was achieved over 90\% efficiency in underwater test \cite{Agostinho}. Yan \emph{et al.} proposed a rotation-free wireless AUVs charging system based on a new coil structure to achieve stable output power. The new coil structure has two decoupled receivers, consisting of two reverse wound receiver coils with the  magnetic flux directions of the two receivers perpendicular to each other, and can achieve 92\% DC-DC efficiency\cite{Y_Zheng}. An ultra-fast inductive power transfer system for AUVs charging which based on a hull compatible coil structure is proposed by Mostafa \emph{et al.} The system can achieve 96.8\% efficiency in saltwater \cite{Mostafa}. Wu \emph{et al.} presented a docking design based on a funnel docking station with an acoustic and optical combination guiding method and a position locking mechanism, which was proposed to enable smooth docking of the AUVs and to eliminate dynamic disturbance during charging. A novel magnetic structure with an overlapped direct-quadrature (D-Q) transmitter and dipole receivers is also proposed to eliminate dynamic disturbance during charging \cite{Sh_Wu}. Cai \emph{et al.} proposes a cross-coupled magnetic coupler that has the advantage of being surface adapted and receiver light, and supports stable and efficient charging of AUVs. A prototype was built and tested on a 600 g receiver which could deliver 1 kW of power at 95.1\% DC-DC efficiency of 95\cite{CW_Cai}. Zeng \emph{et al.} presented a new hybrid transmitter consisting of conical and planar spiral coils which greatly improved the misalignment tolerance and transmission performance of wireless AUVs charging system. The power transfer efficiency was maintained at around 86\% in the excessive misalignment area with transmission distance of 2cm \cite{Y_Zeng}.

\begin{figure}
	\begin{center}
		\includegraphics[width=0.5\textwidth]{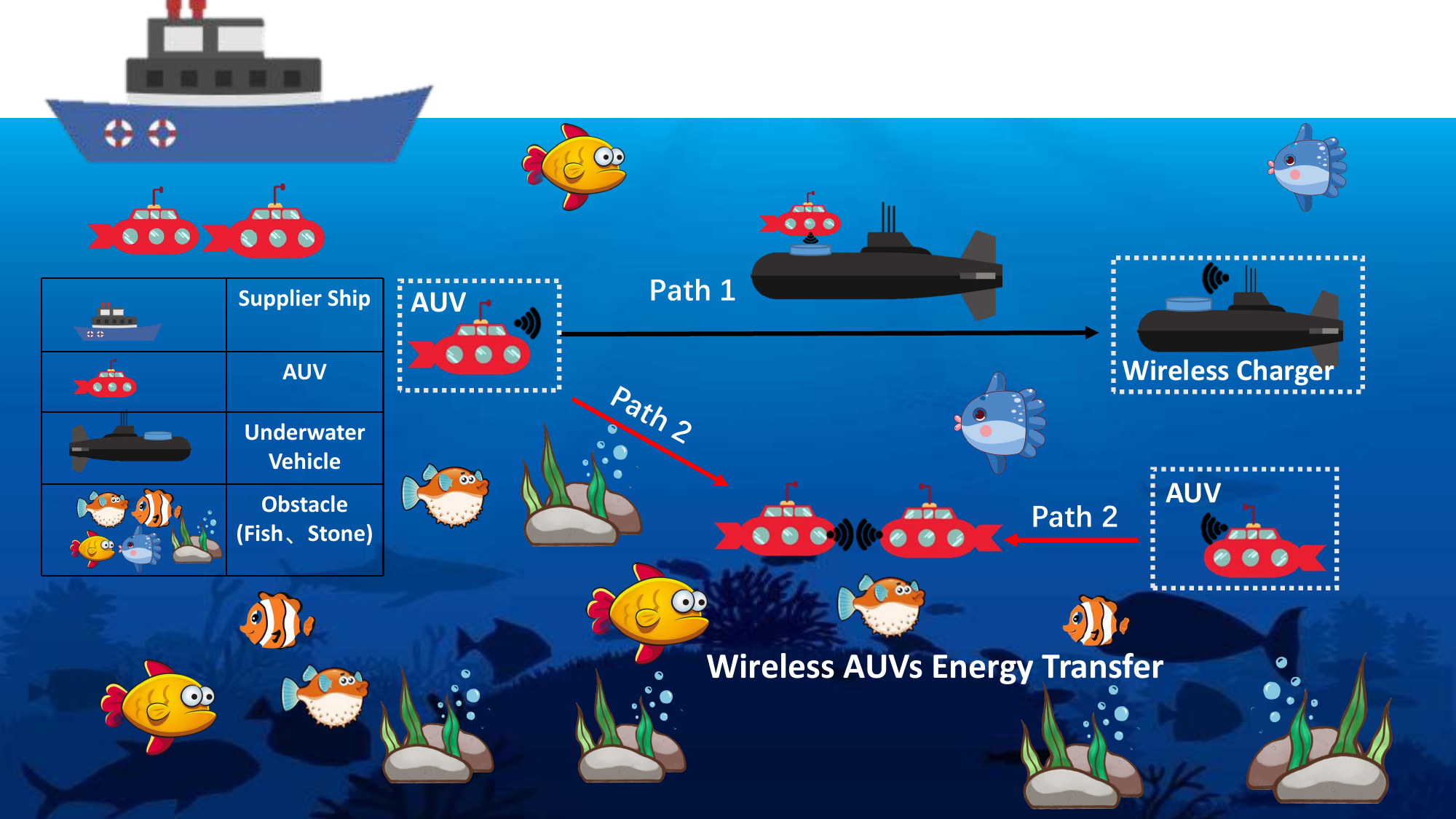}	
	\end{center}
	\caption{The Framework of Wireless AUVs Energy Sharing System}
	\label{Fig.1}
\end{figure}

In conventional wireless power transfer technology the transmitter needs to be fixed, however, the underwater environment is complex and it is difficult to fix the transmitter. Moreover, the AUVs' activity scenarios may be to be in the middle of the ocean, rather than on the sea floor, where the transmitter is not able to be fixed. In this paper, a framework of the wireless AUVs energy sharing system is proposed. When the target consumer AUV out of power, it can send a charging request through the onboard system and select the appropriate AUV from the interface and choose the charging location to complete the energy sharing and charging demand. Fig.~\ref{Fig.1} displays the framework of the wireless AUVs energy sharing system. When the target consumer AUV sends a charging request, the onboard system of other AUV around the target consumer AUV automatically positioning the available supplier AUV and calculates the available energy, distance, AUVs type, \emph{etc.} After this procedure, it returns this information to the target consumer AUV. The target consumer AUV can select the appropriate AUV from the interface and choose the charging location to complete the charging demand. A very important aspect of the wireless AUVs energy sharing system is the path selection after matching the target consumer AUV with the supplier AUV. Due to the complexity of the underwater environment, line of sight obstructions and the presence of static or random obstacles such as coral reefs and schools of fish.

With the improvement of AUV's information acquisition ability, AUVs are able to voyage from one place to another guided by a global reference path. Path planning is a crucial aspect of autonomous voyage, integrating information gathered by sensors and then making choice according to algorithm. A number of algorithms have been proposed in order to generate safe paths efficiency. Depending on the type of configuration space modelling, path planning algorithm can be divided into the sampling-based algorithms and the grid-based algorithms. The sampling-based algorithms can be divided into single-query path planning and asymptotically optimal path planning. The former focuses on obtaining an feasible path quickly without concerning the path length, such as Probabilistic Road Map (PRM)\cite{J_Xu}, Rapid Exploration Random Tree(RRT)\cite{L_Juan} and it is variants RRT-connect\cite{Li_Shuyu}, Lazy RRT. The latter prefer to obtain the optimal path, while may consume more time, such as DT-RRT*\cite{Chen_Long}, Informed RRT*-Connect\cite{Mashayekhi}, Regionally accelerated batch informed trees(RABIT*)\cite{Choudhury}, Hybrid RRT\cite{Mashayekhi_Reza}, these methods generally require a trade-off between path length and time complexity due to the stepwise optimization feature. However, sampling-based algorithms do not require discretization of the environment, which increases the spatial complexity of algorithm space. Grid-based methods such as A*\cite{Liu_X}, D*, Dijkstra\cite{Gbadamosi} and genetic heuristic algorithm such as particle swarm optimisation (PSO) and other genetic algorithms\cite{Li_Lin}. All of the above mentioned are global path planning, and as the scale of the problem expend, the worse the optimisation efficiency and path length becomes. In addition, there are many researches focusing on local path planning, such as Dynamic Window Approach (DWA)\cite{Shen_Yue}, potential field methods such as Artificial Potential Field(APF)\cite{Keyu} and Interfered Fluid Dynamical System(IFDS)\cite{Jianfa_Wu}. These methods have the advantage of low computational complexity and good real-time performance in dynamic environments, but they have proved to have the unavoidable disadvantage of being prone to falling into local minima and, in particular, to oscillations when irregular obstacles are encountered.

In this article, RRT* and PSO are used to test the real-time and optimal routing performance of AUVs path planning, and compare RRT* and PSO in terms of time, path length and search space. Section $\mathrm{II}$ and section $\mathrm{III}$ refer to the principles of RRT* and PSO and propose ideas for path planning based on these two algorithms for modelling AUVs respectively.  Section $\mathrm{IV}$ is the simulation results and discussion of RRT* and PSO in different environments. Eventually section $\mathrm{V}$ is the conclusion.

\section{Rapidly-exploring Random Trees Star}
\begin{figure}[htbp]
	\begin{center}
		\includegraphics[width=1.1\linewidth]{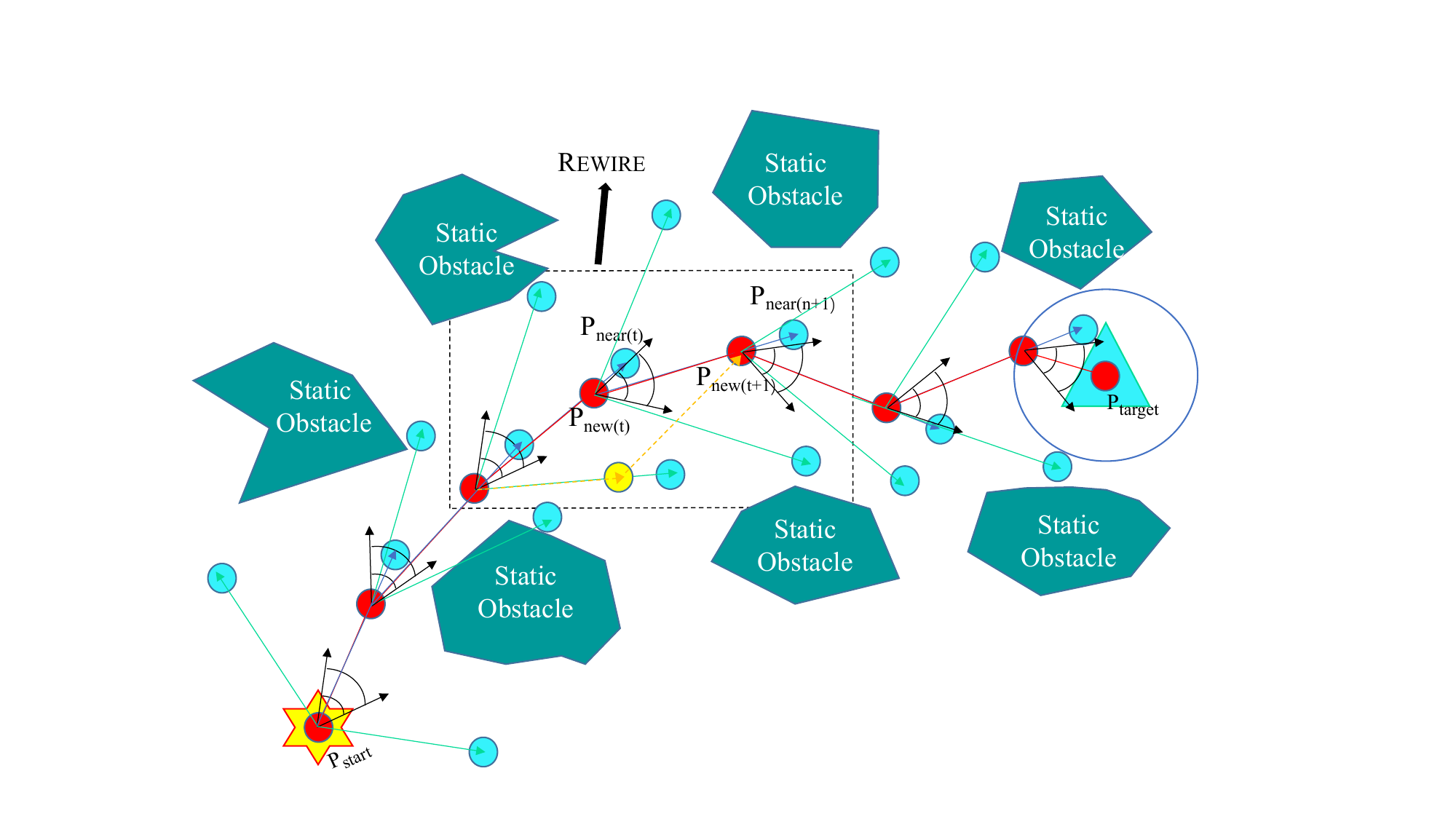}
	\end{center}
	\caption{The Model of AUVs' Trajectory in RRT*}
	\label{Fig.2}
\end{figure}
\subsection{Problem Formulation}
Denoting $\mathbf{X} \in \mathbb{R}^3$ to be the UAV's ROM (Range Of Motion), $\mathbf{X}_{\mathrm{obs}}$ as the obstacles region and $\mathbf{X} \setminus \mathbf{X}_{\mathrm{obs}}$ the open set, and the obstacles-free space as $\mathbf{X}_{\mathrm{free}}=cl(\mathbf{X}\setminus\mathbf{X}_{\mathrm{obs}})$\cite{Dam_T}. All points in $\mathbf{X}_{\mathrm{free}}$ can be selected, but points in $\mathbf{X}_{\mathrm{OBS}}$ are prohibit. The start point is denoted as $\mathrm{P_{start}}=(x_s,y_s)$, the goal point is $\mathrm{P_{target}}=(x_e,y_e)$. The path planning problem is defined to obtain a path $\tau:[0,N]\rightarrow \mathbf{X}_{\mathrm{free}}$ while $\tau(0)=\mathrm{P_{start}}$ and $\tau(N)=\mathrm{P_{target}}$. Define $\sum$ as the set of all feasible paths. $C(\tau)$ is the cost function of each paths $\tau$. The path planning problem is defined as a trajectory $\tau$ from initial place to goal place, meanwhile avoiding obstacles and ensure $C(\tau)$ as small as possible. In this paper, we use RRT* and PSO to achieve path planning and compare the time, length and search space of the two.

\subsection{Informed Points and Modeling}
The model of AUVs' trajectory in RRT* is illustrated in fig.~\ref{Fig.2}. A potential solution of path planing is defined:

 $\tau =\left\{ \mathrm{P_{start}},\mathrm{P_{new(1)}}, \mathrm{P_{new(2)}},..., \mathrm{P_{new(t)}, \mathrm{P_{new(t+1)}}, \mathrm{P_{target}}}\right\}$. 
 Small blue dots are the sampling points generated by each iteration, small red dots are the selected points, the small yellow dot is the selected point but removed in the $\mathrm{Rewire}$ process, and the big blue circle is the range circle of target point. The planned path is shown in the red solid line. 

\subsection{RRT* Algorithm}
The RRT* algorithm in this paper is designed as shown in Alg.~\ref{Alg}. Ensuring the AUV's safe navigation in the underwater environment is critical. Dynamic obstacles like fish, constant obstacles like reefs in the seafloor should be avoided. In this paper, Euclidean Space is considered, and the cost of the path stiffness distance between two points is the positive Euclidean distance between two points, which is expressed as
\begin{equation}
	Dist\left\{P_{new(t)}, P_{new(t+1)} \right\} = |P_{new(t)}- P_{new(t+1)}|
	\label{E}
\end{equation}

Starting point $P_{start}$, target point $P_{target}$, spatial obstacle information $Obstacles$ and cartographic information $Map$ are defined. The whole space is initialized with the sampling points $IterationsNum$. Initialize the entire space, and set the sampling number $IterationsNum$. $StepSize$ between points, and minimum distance threshold $MinThreshold$. The algorithm will start its iterative process, firstly it generates a point $P_{rand}$ randomly in the space, then finds the point $P_{near}$ which is closest to this random point in the point set of known tree, obtains point $P_{new}$ with $StepSize$ in the direction from $P_{near}$ to $P_{rand}$, and judges whether there are obstacles between $P_{near}$ and $P_{new}$. If it exists, discard the $P_{new}$, if no, it opens $\mathbf{GetNeighbors}$ to obtain the new node index. If there is no index, insert $P_{near}$ into the tree to check whether $P_{neighbors}$ exists and update again. Continue the cycle until $P_{new}$ is in the specified neighborhood of the destination, obtain the shortest distance. The algorithm is then finished.
\begin{algorithm}[h]
	\caption{Rapidly-exploring Random Trees Star Algorithm}
	\label{Alg}	
	\KwIn{$P_{start}$,  $P_{target}$, $Obstacles$, $Map$ }
	\KwOut{A Path $\tau$ in $RRT*$ from $P_{start}$ to $P_{goal}$}
	%ensure $ optimal path from $P_s$ to $P_g$
	$RRT*$.$\mathbf{Init}$($IterationsNum$, $StepSize$, $MinThreshold$)\;
	\While{$i <= IterationsNum$}{\
		$P_{random} \leftarrow \mathbf{RandomSample}$($Map$)\;\
		$P_{near} \leftarrow \mathbf{FindNearest}$($RRT*$, $P_{rand}$)\;\
		$P_{new} \leftarrow \mathbf{Steering}$($P_{random}$, $P_{near}$, $StepSize$)\; \
		\If{$\mathbf{ObstaclesDetection}$($Obstacles$, $P_{new}$)}
		{ 
			$P_{neighbors} \leftarrow \mathbf {GetNeighbors}$($RRT*$, $P_{new}$)\;
			\eIf{$\mathbf{EmptyJudging}$($P_{neighbors}$)}{
				$P_{parent} \leftarrow \mathbf{ParentChoosing}$($RRT*$, $P_{neighbors}$, $P_{near}$, $P_{new}$)\;
			}
			{$P_{parent} \leftarrow P_{near}$\;}
			${RRT*} \leftarrow \mathbf{NodeInsert}$($RRT*$, $P_{parent}$, $P_{new}$)\;
			\If{$\mathbf{EmptyJudging}$($P_{neighbors}$)}{
				$RRT* \leftarrow \mathbf{RewireRRT*}$($RRT*$, $P_{neighbors}$, $P_{parent}$, $\mathbf{Length}$($RRT*$))\;
			}
		}
	}
	\If{$\mathbf {Distance}$($P_{new}$, $P_{target}$)$ <= MinThreshold$}{
		$\tau \leftarrow \mathbf{GetOptimizedPath}$($RRT*$)\;
		return $\tau$\;
	}	
\end{algorithm}

\section{Particle Swarm Optimization}

\begin{figure}[htbp]
	\begin{center}
		\includegraphics[width=1.1\linewidth]{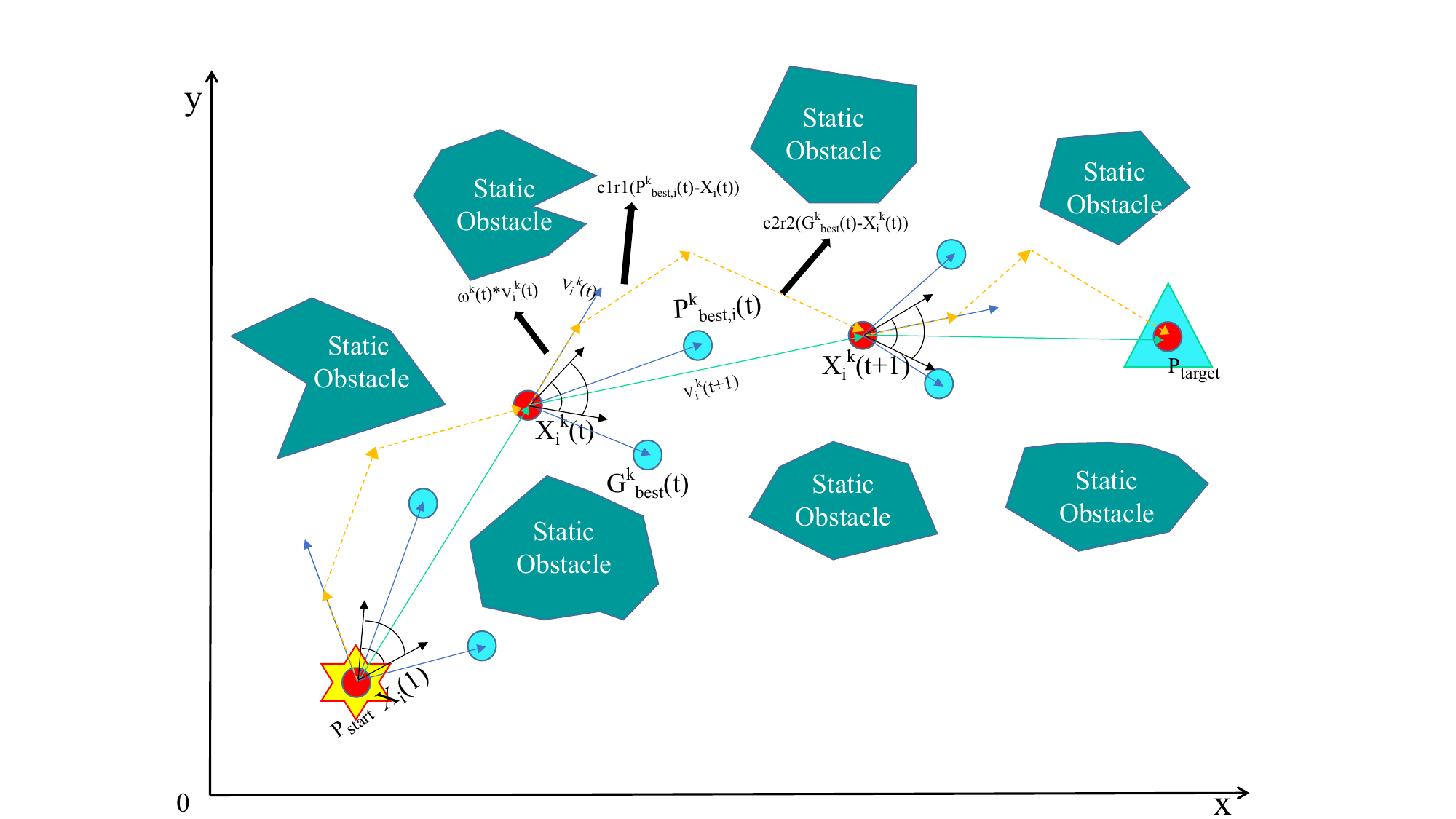}
	\end{center}
	\caption{The Model of AUVs' Trajectory in PSO}
	\label{Fig.3}
\end{figure}

\subsection{Formulas and Symbols}
In this paper, we define The velocity and position of particle i at time t as $V_{i}(t)$, $X_{i}(t)$, the individual learning factor of particle i, generation k at time t as $P_{best,i}^k(t)$ and the social learning factor of particle i, generation k as $G_{best}^k(t)$, the formula of velocity update $V_{i}^k(t)$ and position update $X_{i}^k(t)$ are as follows:
\begin{equation}%加*表示不对公式编号
	\begin{split}
		V_{i}^k(t+1)=&\omega^k(t) V_{i}^k(t)+c_{1}r_{1}(P_{best,i}^k(t)-X_{i}^k(t))\\
		&+c_{2}r_{2}(G_{best}^k(t)-X_{i}^k(t)) 
	\end{split}
\end{equation}
\begin{equation}
	\begin{split}
		X_{i}^k(t+1)=X_{i}^k(t)+V_{i}^{k}(t+1)
	\end{split}
\end{equation}

As shown in Fig.~\ref{Fig.3}, in the Cartesian coordinate system, the connection of each two small red dots is the direction of motion, and the small blue dot is one of individual optimal and group optimal. The solution path is obtained by linear addition of the line vector of the position of small red and blue dots and the velocity inertia of the red particle itself.
%In this paper,the individual optimization function and the population optimization function are defined as $f_{p}$ and $f_{y}$.

\subsection{PSO Algorithm}
The steps of PSO algorithm to solve path planning are as follows and the flow chart is shown in Fig.~\ref{Fig.4}. 
\begin{enumerate}
	\item {Explore and model the environment, initialize parameters, enter the coordinates of start and target.}
	\item {Initialize the position and speed of each particle in the particle swarm, and set Maximum Number of Iterations (MI) and Population Size (nP).In the process of updating the position of particles, a section of the route will touch the obstacle. In the program, punishment is imposed to drive the route to avoid the obstacle.}
	\item {The main loop starts iterating, and updates the speed and position of the particle every time. Take the speed and position of the particle of t as an example: \\
    Update the current particle velocity, the current velocity boundary value, the current particle position, current particle position boundary value in the x direction and the y direction, respectively.\\
	Update the individual optimal particle and global optimal particle, evaluate the objective function value of the particle, and update the inertia weight of its own motion.\\
	Until the maximum number of iteration steps is reached or an acceptable solution is obtained: the optimization stops when the difference between the adaptive value of the optimal solution after the last iteration and that of the optimal solution after the current iteration is less than the minimum threshold.} 
	\item {Output the global optimal particle, the optimal particle is the feasible path. }
\end{enumerate}
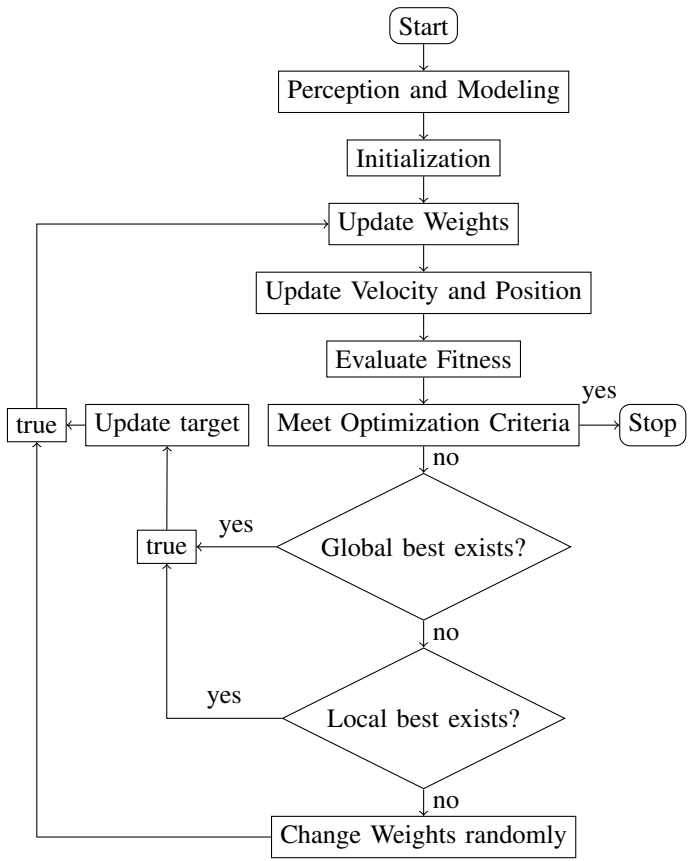
\begin{figure}[htbp]
	\begin{tikzpicture}[node distance=10pt]
		\node[draw, rounded corners]                        (start)   {Start};
		\node[draw, below=of start]      					(Perception and Modeling)  {Perception and Modeling};
		\node[draw, below=of Perception and Modeling]                         (Initialization)  {Initialization};
		\node[draw, below=of Initialization]                        (Update Weights)  {Update Weights};
		\node[draw, below=of Update Weights]    (Update Velocity and Position)  {Update Velocity and Position};
		%\node[draw, left=60pt of Update Weights]                   (true1)  {true};
		\node[draw, below=of Update Velocity and Position]                        (Evaluate Fitness)  {Evaluate Fitness};
		
		\node[draw, below=of Evaluate Fitness]                        (Meet Optimization Criteria?)  {Meet Optimization Criteria};
		\node[draw,rounded corners, right=15pt of Meet Optimization Criteria?]                   (Stop)  {Stop};
		
		\node[draw, left=7pt of Meet Optimization Criteria?]                   (Update target)  {Update target};
		\node[draw, left=7pt of Update target]                   (true1)  {true};
		\node[draw, diamond, aspect=2,below=of Meet Optimization Criteria?]                        (Global best exists?)  {Global best exists?};
		\node[draw, diamond, aspect=2,below=of Global best exists?]                        (Local best exists?)  {Local best exists?};
		\node[draw, left=30pt of Global best exists?]                  (true2)  {true};
		\node[draw, below=of Local best exists?]                        (Change Weights randomly)  {Change Weights randomly};
		
		\draw[->] (Meet Optimization Criteria?) -- node[above=4.5pt]    {yes}(Stop);
		\draw[->] (start)  -- (Perception and Modeling);
		\draw[->] (Perception and Modeling)  -- (Initialization);
		\draw[->] (Initialization) -- (Update Weights);
		\draw[->] (Update Weights) -- (Update Velocity and Position);
		\draw[->] (Update Velocity and Position) -- (Evaluate Fitness);
		\draw[->] (Evaluate Fitness) -- (Meet Optimization Criteria?);
		\draw[->] (Meet Optimization Criteria?) --node[right] {no} (Global best exists?);
		\draw[->] (Global best exists?) --node[right] {no}(Local best exists?);
		\draw[->] (true2) -- (Update target);
		\draw[->] (Local best exists?) -- node[right] {no}(Change Weights randomly);
		
		%\draw[->] (true)  -- (Update Weights);
		
		%\draw[->] (Change Weights randomly) -- (Update Weights);
		\draw[->] (Global best exists?) --node[above] {yes} (true2);
		\draw[->] (Local best exists?) --node[above] {yes} (Local best exists?-|true2) -> (true2);
		\draw[->] (Update target)  --  (true1);
		\draw[->] (Change Weights randomly)  -|  (true1);
		\draw[->] (true1)  |-  (Update Weights);
		
	\end{tikzpicture}\
	\caption{The Flow Chart of PSO}
	\label{Fig.4}
\end{figure}
\section{Simulation and Results}
In this section, the performance of the algorithm will be evaluated in MATLAB2020b on a computer with an Intel Core i5-1135G7 at 2.4GHz and 16GB RAM. In order to compare and analyze the performance of $\mathrm{RRT*}$ and $\mathrm{PSO}$ in the same situation. One is a general random obstacle, with the aim of testing the performance of algorithm in most cases. The second is a specific irregular obstacles which is designed to test corner cases in order to compare the robustness of the algorithms.
\begin{table}[h]
	\renewcommand\arraystretch{1.5}
	\tabcolsep=0.25cm
	\caption{Path Length Distribution in Ten Cases}
	\label{table1}
	\begin{tabular}{lllllll}
		
		\toprule[1pt]
		\multicolumn{1}{r}{Case} & $P{start}$ & ${target}$ & RRT* & Exist? & PSO & Exist? \\ \hline
1 & (12,-35) & (-15,10) &\textbf{67.730} & \textbf{    yes} & \textbf{56.008} & \textbf{	yes} \\
2 & (10,-30) & (-20,8) & \textbf{63.083} & \textbf{    yes} & \textbf{49.761} & \textbf{	yes} \\
3 & (25,-35) & (-7,10) & \textbf{63.616} & \textbf{	   no} & \textbf{71.003} & \textbf{     yes} \\
4 & (5,-28) & (10,13) & \textbf{54.242} & \textbf{    yes} & \textbf{82.873} & \textbf{   yes} \\
5 & (0,-32) & (12,10) & \textbf{56.522} & \textbf{    yes} & \textbf{153.304} & \textbf{	no} \\
6 & (-2,-33) & (13,15) & \textbf{68.350} & \textbf{    no} & \textbf{129.371} & \textbf{	no} \\
7 & (-10,-30) & (20,10) & \textbf{61.072} & \textbf{    yes} & \textbf{87.124} & \textbf{	yes} \\
8 & (25,8) & (-12,-25) & \textbf{64.059} & \textbf{    no} & \textbf{53.319} & \textbf{	no} \\
9 & (-38,-10) & (32,-10) & \textbf{85.123} & \textbf{    yes} & \textbf{82.062} & \textbf{	yes} \\
10 & (33,-7) & (-20,-13) & \textbf{87.848} & \textbf{    yes} & \textbf{55.921} & \textbf{		yes}\\
		\bottomrule[1pt]
	\end{tabular}
\end{table}
\subsection{Multi-Random Obstacles}\label{AA}

\begin{figure}[htbp]
	\centering
	\subfigure[RRT*]{
		\includegraphics[width=0.5\textwidth,height=0.3\textwidth]{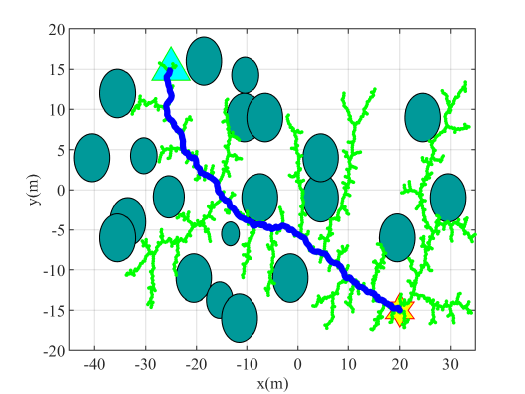}
	}	
	\subfigure[PSO]{
		\includegraphics[width=0.5\textwidth,height=0.3\textwidth]{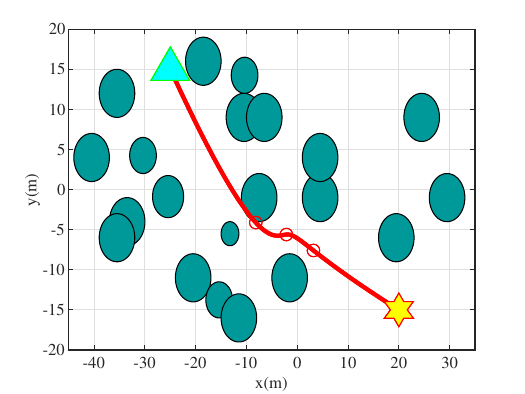}
	}
	\caption{The Path Result of Two Algorithms in Random Obstacles}
	\label{Fig.5}
\end{figure}

In this environment, The RRT*, PSO were tested to show their ability to find optimal path while avoiding multi-random obstacles. To make the environment fair, the maximum number of iterations for three algorithms was equal to $2000$. However, if the algorithm gets solution before maximum iteration, it will also be counted. The starting point is $P_{start}=(20,-15)$, and the target point is $P_{target}=(-25,15)$. The results are shown in Fig.~\ref{Fig.5}. To reveal the average level of algorithm performance in random obstacles, the experiments will be run 50 times and the running time results of two algorithms are shown in Fig.~\ref{Fig.5}.

\begin{figure}[htbp]
	\begin{center}
		\includegraphics[width=1.1\linewidth]{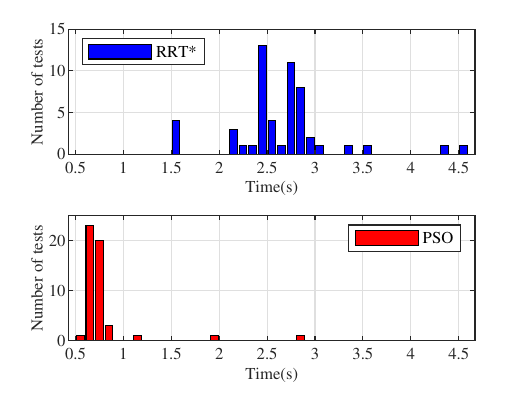}
	\end{center}
	\caption{The Comparison of Time Distribution.}
	\label{Fig.6}
\end{figure}

\begin{figure}[htbp]
	\begin{center}
		\includegraphics[width=0.5\textwidth,height=0.35\textwidth]{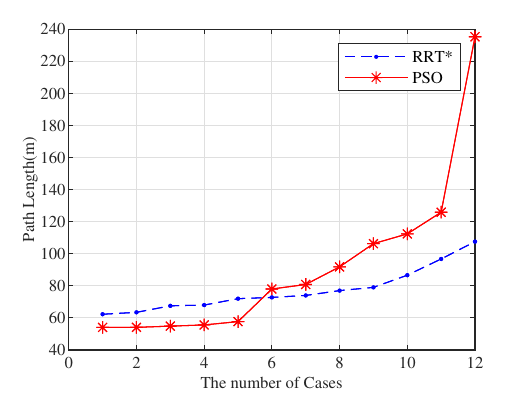}
	\end{center}
	\caption{The comparison of Path Length Distribution.}
	\label{Fig.7}
\end{figure}

By analyzing the Fig. ~\ref{Fig.5}, it is found that they both find solutions while avoiding obstacles. Fig.~\ref{Fig.6} shows the distribution of the two path lengths. It can be seen that the path length distribution of PSO is very unstable, with a standard deviation of $52.6129$ and an average path length of $97.7755$. The standard deviation of RRT* is $13.0681$, the average path length is $77.6423$.

By evaluating the index time and path length, it can be seen that RRT* does a lot of useless work throughout the spatial search as it is a random search, whereas PSO cannot minimize the path due to the limitation of maximum iterations. the average path length of PSO is longer than that of RRT*, but the time consumption of 50 iterations is better distributed. As shown in Fig.~\ref{Fig.7}, RRT * time mainly for $[2.1, 3.0]$, PSO time mainly for $[0.6, 0.8]$. In general, PSO is not guaranteed to find the globally optimal path. In terms of path length, PSO is worse than RRT*, but in terms of time, it is better than RRT*.

\begin{figure}[h]
	\centering
	\subfigure[RRT*]{
		\includegraphics[width=0.5\textwidth,height=0.35\textwidth]{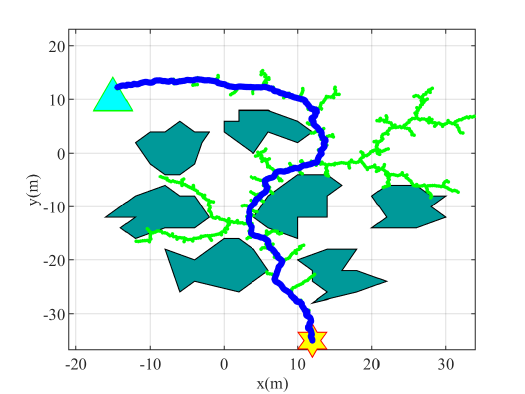}
	}
	
	\subfigure[PSO]{
		\includegraphics[width=0.5\textwidth,height=0.35\textwidth]{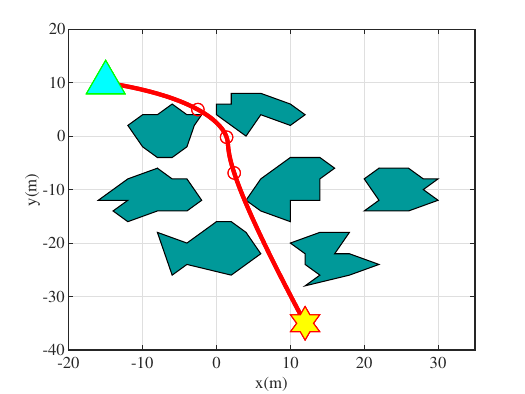}
	}
	\caption{The Path Results of two algorithms in Specific Irregular Obstacles.}
	\label{Fig.8}
\end{figure}

\subsection{Specific irregular obstacles}

In this section, RRT*, PSO are tested in specific irregular obstacles which are more similar to the real situation. As shown in Fig.\ref{Fig.8}, the starting point is $[12,-35]$ and the target point is $[-15,10]$. The results of RRT* show a large number of small angle turns, which have a poor effect on the smoothness of the path and the overall path length. In contrast, PSO have a smoother path than RRT*. In addition, the search space of PSO is smaller than RRT*, so PSO is more efficient than RRT*. As shown in Table.\ref{table1}, set different initial points and target points, and analyze the path length of the two algorithms, both the PSO and the RRT* crossed obstacles or failed to reach the target point three times, RRT* algorithm is due to the fact that the path cannot be planned normally when the exploration space is narrow, and the path length is shorter if it cannot pass, Because the number of iterations is limited, the PSO is not optimal, and it shows a longer length of road stiffness. In a word, the path length of PSO is shorter than RRT* under normal path planning.

\section{Conclusion}

In this article, we proposed a framework of wireless AUVs energy sharing system, which can provide fast energy rescue for AUVs and extend their activity range. Path planning is an important part of the system, especially the matching between consumer AUVs and supplier AUVs. As for the two algorithms summarized in this paper, the path generated by PSO takes less time, has a higher smoothness of road force, avoids random, irregular obstacles, consumes less energy, and makes the AUV's navigation from the initial position to the target position more disciplined. However, the disadvantage of PSO is that it cannot guarantee convergence to the global optimal point to get the global optimal path. RRT* selects the parent node again on the basis of the RRT Algorithm to minimize the path cost of the newly generated node. Rewiring makes the random tree after the generation of new nodes reduce redundant paths, reduce the path cost, optimize the path point selection, and improve the algorithm efficiency. However, the underwater vehicle will be affected by the current flow in the Marine environment, so future work will focus on more suitable for the real Marine environment and smooth path curvature, making the underwater vehicle navigation more satisfying to the motion model.


%\bibliography{reference}
\bibliographystyle{IEEEtran}

\end{document}